# Analysis and optimization of a novel energy storage flywheel for improved energy capacity.

Xiaojun Li[a,b], Alan Palazzolo[a]

[a]*Dwight Look College of Engineering, Texas A&M University, College Station, TX, 77840, USA*
[b]*Gotion Inc, Fremont, CA, 94538, USA*

**Abstract**

Kinetic/Flywheel energy storage systems (FESS) have re-emerged as a vital technology in many areas such as smart grid, renewable energy, electric vehicle, and high-power applications. FESSs are designed and optimized to have higher energy per mass (specific energy) and volume (energy density). Prior research, such as the use of high-strength materials and the reduction of stress concentration, primarily focused on designing and optimizing the rotor itself. However, a modern FESS includes other indispensable components such as magnetic bearings and a motor/generator that requires a shaft. The shaft significantly impacts the flywheel design. This paper investigates several typical flywheel designs and their stress analysis. A simplified analysis method is given for designing rotor-shaft assembly. It is found that the shaftless flywheel design approach can double the energy density level when compared to typical designs. The shaftless flywheel is further optimized using finite element analysis with the magnetic bearing and motor/generators' design considerations.

Keywords: Battery, Energy storage flywheel, Shaft-less flywheel, Renewable energy, Stress analysis, Design optimization

# Introduction

As one of the alternatives to lithium-ion batteries [1], the FESS technology has been increasingly commercialized and applied to different areas[2,3]. As one of the early pioneers, Beacon Power Corporation commissioned a frequency regulation power plant with flywheels. The project costs over 40 million dollars and has a 20MW peak power output [4]. Based on estimations, a single unit costs around 260k and can store 25KWh[5]. The flywheel consists of a composite rotor/rim with a metallic shaft, with a max spinning speed of 16,000 RPM. More recently. The Calnetix/Vycon VDC/REGEN system [6] is commercially targeted at mission-critical applications such as hospitals and data centers. The REGEN model has been successfully applied to the L.A. metro subway [7] as a Wayside Energy Storage Substation (WESS). It was reported that the system had saved $10-18 worth of traction energy daily. The analysis in [7] shows that "WESS will save at least $99,000 per year ".

Flywheel design is usually the starting point of the system design. Most of the previous research work focuses on the optimization of composite flywheels. Arvin and Bakis [8] proposed a concept design of a flywheel made of multiple rims, press-fit, filament wound composite materials. The optimization uses a simulated annealing algorithm and gives specific energy of 40-50 kWh/kg. Abrahamsson et al. [9] presented an optimized high-speed kinetic buffer flywheel. The rotor comprises a solid composite shell of carbon and glass fibers in an epoxy matrix constructed in one curing. Ha et al. [10] studied the effects of interferences and rim thickness to maximize the specific energy of a multi-rim composite flywheel rotor. Up to five rims of graphite/Ep and glass/Ep are analyzed. Others have studied the effects of flywheel's geometric profile. Arslan [11] used finite element analysis to examine 6 cases with different geometric profiles. It was found that the best case gives specific energy of 8.977 Wh/ kg. Kress [12] used a 2D finite element model to optimize a bored flywheel.

The kinetic energy ($E$) stored in a flywheel is given by

$$E = \frac{1}{2} I_p \omega^2 \qquad (1)$$

where $I_p$ is the moment of inertia, and $\omega$ is the flywheel spinning speed. Flywheels are designed to have a higher moment of inertia and rotate at a higher spinning speed to raise the energy capacity. High spinning speed eventually leads to a failure caused by the stress developed by inertia load [13]. Genta has laid out the foundation of flywheel designs by researching theoretical analysis of isotropic flywheels. The following equation (2) is introduced to characterize the factors that determine a flywheel's specific energy:

$$\frac{E}{m} = K \frac{\sigma}{\rho} \qquad (2)$$

where K is the shape factor, $\sigma$ the material's tensile strength, and $\rho$ the density.

Composite materials are often chosen to make FESS flywheels for low density and high tensile strength. They may have a very high specific energy, crucial in aerospace or mobile applications. Research works in [10,14,15] have claimed very high specific energies that reach 50 to 100 Wh/kg. However, only the composite rim was included in the calculation. The metallic shaft, an essential component with considerable mass, is usually neglected. Total energy density is reduced when considering the shaft as well. For example, one of the composites-based FESS being successfully developed [16] has a specific energy of 42KJ/kg, equivalent to only 11.7Wh/kg. The specific energy drops to 5.6Wh/kg when the entire system weight is considered. The cost of composite materials is significantly higher than steel. Based on [13], the comparisons of density, tensile strength, and costs between composite and steel are summarized in Table I . While Carbon

T1000 may have a density of only 20% of steel's and tensile strength 26% higher than steel, its cost is almost 100 times more. Notice that the maximum specific energy of T1000 is based on an ideal and pure composite flywheel. The energy capacity per dollar of composite flywheels is only a fraction of steel flywheels. From the cost standpoint, steel-based FESSs are more suitable for massive production. Large-scale, ground-based FESS applications are less reliant on system weight reduction. High Strength Steel (HSS) flywheels have a high energy density (volume-based energy) due to their high mass density, therefore they are very suitable for fixed, ground-based, large capacity, widespread applications. Furthermore, HSS flywheels are superior to composite ones in terms of thermal conductivity and availability of design data such as S.N. curves, fracture toughness, etc.

Table I Comparison of different rotor materials [13]

| Materials | Density (kg/m3) | Tensile strength (MPa) | Max specific energy (Wh/kg) | cost ($/kg) | Energy per dollar (Wh/$) |
|---|---|---|---|---|---|
| Steel 4340 | 7700 | 1520 | 50 | 1 | 50 |
| E-glass | 2000 | 100 | 14 | 11 | 1.27 |
| S2-glass | 1920 | 1470 | 210 | 24.6 | 8.54 |
| Carbon T1000 | 1520 | 1950 | 350 | 101.8 | 3.47 |
| Carbon AS4C | 1510 | 1650 | 300 | 31.3 | 9.59 |

FESSs are complex systems that require integrating mechanical, magnetic, and electrical systems. Specific energy determines the rotor's cost, which eventually drives the overall cost of the entire FESS. Thus, the design and analysis of the rotor is usually the starting point of building a FESS. On the other hand, the significance of other components must be considered during the initial design stage. Texas A&M University has developed a shaftless flywheel energy storage system [17,18] with a coreless motor/generator [19]. The system is aimed at:

1. To minimize the cost per kWh and kW for FESSs
2. To increase the recyclability and reduce the environmental impact of FESSs

In the remainder of this paper, we first propose a simplified flywheel design criterion, considering rotor-shaft assembly. Secondly, different flywheel designs' stress distribution and specific energy are formulated and compared to the shaftless flywheel. In the last section, the detailed shaftless design is briefly introduced. An improvement that utilizes preloading with a new radial ring, together with parametric study, significantly reduces the stress caused by rotation. A new flywheel design with higher specific energy is achieved.

## Stress Analysis

This chapter first discusses the basic stress analysis for energy storage flywheels, including the stress caused by flywheel rotation and external pressures. Then a new stress analysis formula is introduced as a simplified design criterion for shaft-rotor assemble.

### Basic Analysis

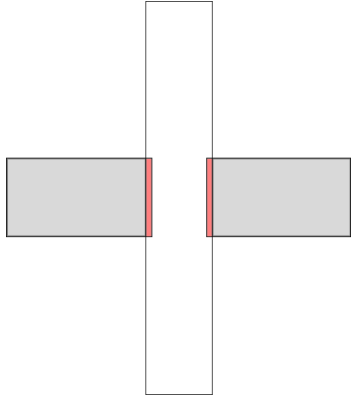

Figure 1 A typical steel flywheel configuration, including an annulus and a center shaft

During the rotational motion, a flywheel is subject to the hoop ($\sigma_\theta$) and radial ($\sigma_r$) stress but has only the radial displacement $u(r)$. The equilibrium equation [20] of a rotating disk is given in (3):

$$\frac{d}{dr}(r\sigma_r) - \sigma_\theta + \rho\omega^2 r^2 = 0 \qquad (3)$$

Von-Mises is widely used as the failure criteria:

$$\sigma_v = \sqrt{\sigma_r^2 + \sigma_\theta^2 - \sigma_r \sigma_\theta} \qquad (4)$$

The solutions of (3) under two different boundary conditions, based on whether it is annulus or shaftless flywheel, are given in the following [20]:

1. The radial stress ($\sigma_r$) and tangential stress ($\sigma_\theta$) of a rotating solid disk with external pressure imposed on the outer radius:

$$\sigma_r(r) = -p_b + \frac{3+v}{8}\rho\omega^2 b^2 \left(1 - \frac{r^2}{b^2}\right)$$
$$\sigma_\theta(r) = -p_b + \frac{3+v}{8}\rho\omega^2 b^2 \left(1 - \left(\frac{1+3v}{3+v}\right)\frac{r^2}{b^2}\right)$$
(5)

where $a, b$ represents the inner and outer radius of the disk, $\rho$ the material density, $\omega$ the rotational speed, and $v$ the Poisson ratio. The equations include two terms. Compressive stress is caused by and proportional to external pressure $p_b$. The other tensile stress is proportional to $\rho\omega^2 b^2$ and caused by centrifugal force. These two components are independent of each other.

2. The radial stress and tangential stress of a rotating annulus disk with external pressure imposed on the inner and outer radius:

$$\sigma_r(r) = p_a \frac{t^2}{1-t^2}\left(1 - \frac{b^2}{r^2}\right) - p_b \frac{1}{1-t^2}\left(1 - \frac{a^2}{r^2}\right)$$
$$+ \frac{3+v}{8}\rho\omega^2 b^2 \left(t^2 + 1 - \frac{r^2}{b^2} - \frac{a^2}{r^2}\right)$$
$$\sigma_\theta(r) = p_a \frac{t^2}{1-t^2}\left(1 + \frac{b^2}{r^2}\right) - p_b \frac{1}{1-t^2}\left(1 + \frac{a^2}{r^2}\right)$$
$$+ \frac{3+v}{8}\rho\omega^2 b^2 \left(t^2 + 1 - \frac{1+3v}{3+v}\left(\frac{r^2}{b^2}\right) + \frac{a^2}{r^2}\right)$$
(6)

The radial stresses associated with outer and inner radius pressure are compressive. The hoop stress associated with inner radius pressure is tensile, but the hoop stress associated with outer radius pressure is compressive. The causes of stress include external pressure $(p_a, p_b)$ or the rotational motion $(\rho\omega^2 b^2)$. The stresses are products of the quantities above and dimensionless factors, which only depend on materials and geometries. Figure 2 shows how the stresses are distributed and how the ratio of inner to the outer radius $(t)$ affects the stress distributions for an annulus flywheel.

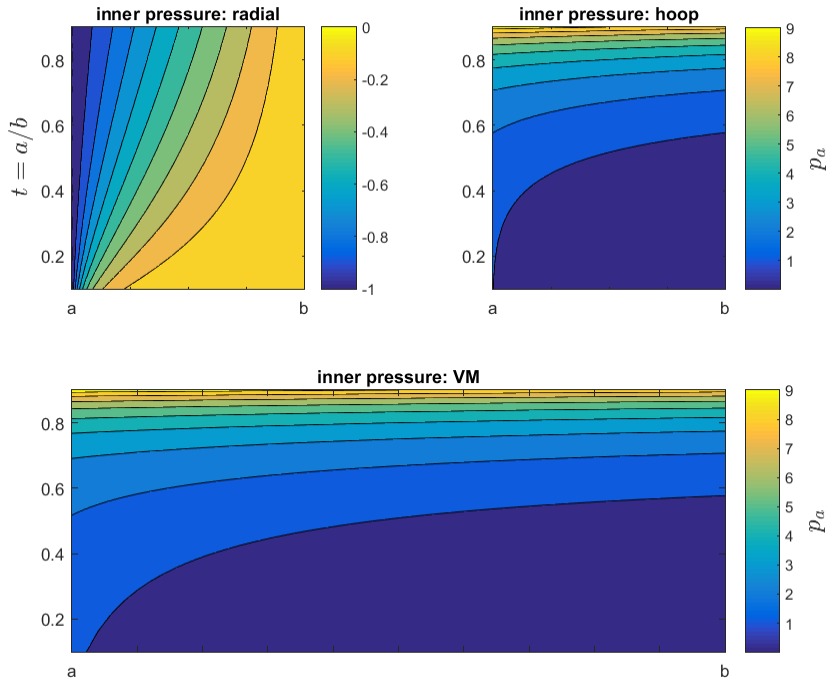

Figure 2 Contour Plot of Inner Radius Pressure Caused Stress: Stress w.r.t. to t (inner to outer radius) and radial position (from a to b) inside an annulus flywheel

Depicted in the contour plots of Figure 2 and Figure 3, radial stresses induced by external pressure (whether inner or outer pressure) are more sensitive to the radial position. In other words, the stress varies significantly with respect to the radial position where it is evaluated. Such variation becomes more linear as $t$ increases. For smaller $t$, stresses are concentrated where the pressure is applied (inner or outer radius). However, increasing $t$ itself has a negligible effect on the maximum value of radial stresses. On the contrary, the inner and outer pressure-induced hoop stress increases significantly when $t$ increases, I.e., when the annulus flywheel becomes thinner, the pressure-induced hoop stress becomes more significant. In addition, stress variation along the radial direction decreases and becomes monotonical as $t$ increases.

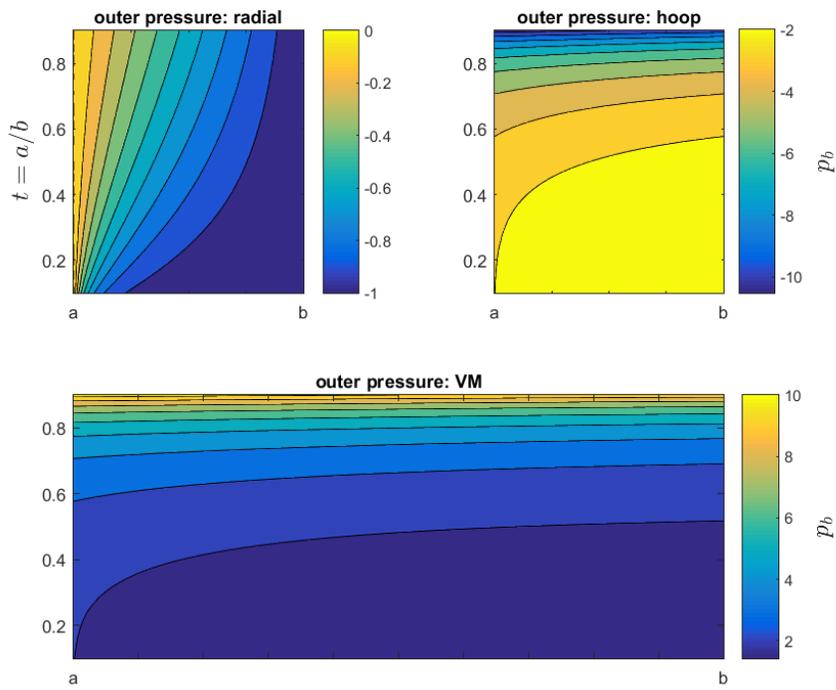

Figure 3 Contour plot of outer radius pressure caused stress w.r.t. to $t$ (inner to outer radius) inside an annulus flywheel.

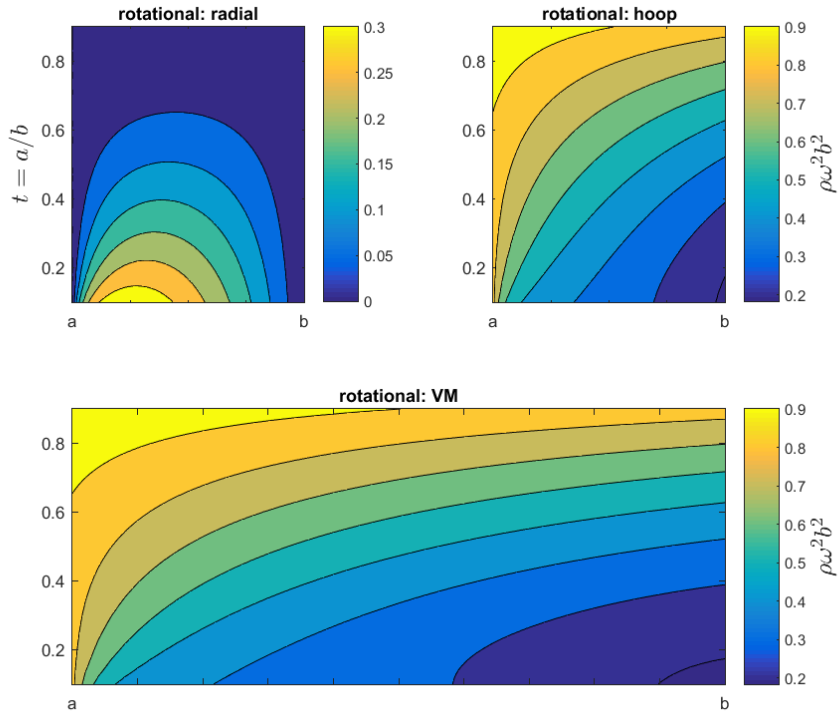

Figure 4 Contour plot of rotational caused stress w.r.t. to $t$ (inner to outer radius) and radial position (from a to b) inside an annulus flywheel.

Rotational induced stresses for an annulus flywheel are given in Figure 4. For the rotational radial stress, the maximum value occurs at $\sqrt{ab}$ along the radial direction. For hoop stress, the maximum value occurs at the inner radius $a$. As the flywheel becomes thinner (as $t$ increases), radial stress decreases but hoop stress increases. For the rotation caused Von-Mises stress, hoop stress is clearly the dominating factor. Because that rotational-induced stress is usually the dominant factor in total stress, the maximum stress of an annulus flywheel would occur at the inner radius.

## Simplified design criteria

Because that the external pressures ($p_a, p_b$) and rotational load ($\rho\omega^2 b^2$) are not directly comparable. To include pressure-induced stress in failure analysis. The following lemma is introduced.

**Lemma**: *The maximum Von-Mises ($\sigma_m$) of an annulus flywheel under shrink-fit pressure and rotational load has a simplified upper boundary, defined as*

$$\sigma_m \leq \sigma_m(\omega) + \sigma_m(p) \tag{7}$$

where $\sigma_m(\omega)$ and $\sigma_m(p)$ are the maximum Von-Mises stress of the flywheel caused by rotation and inner radius pressure.

**Proof:**

$$\sigma_m(\omega + p) = \left.\sqrt{\sigma_r^2 + \sigma_\theta^2 - \sigma_r \sigma_\theta}\right|_{r=a} \tag{8}$$

where $\sigma_r = \sigma_r(p) + \sigma_r(\omega)$, $\sigma_\theta = \sigma_\theta(p) + \sigma_\theta(\omega)$ and $\sigma_r(\omega)|_{r=a} = 0$. Therefore,

$$\sigma_m = \left.\sqrt{\sigma_r^2(p) + \sigma_\theta^2(p) - \sigma_r(p)\sigma_\theta(p) + \sigma_\theta^2(\omega) + 2\sigma_\theta(\omega)\sigma_\theta(p) - \sigma_r(p)\sigma_\theta(\omega)}\right|_{r=a} \tag{9}$$

Since $\sigma_m^2(p) = \sigma_r^2(p) + \sigma_\theta^2(p) - \sigma_r(p)\sigma_\theta(p)$ and $\sigma_m(\omega) = \sigma_\theta(\omega)$

$$\sigma_m = \left.\sqrt{\sigma_m^2(p) + \sigma_m^2(\omega) + 2\sigma_m(\omega)\left(\sigma_\theta(p) - \frac{1}{2}\sigma_r(p)\right)}\right|_{r=a} \tag{10}$$

Since $\sigma_r(p)\sigma_\theta(p) \leq 0$ for inner pressure-caused stress.

$$\left.\left(\sigma_\theta(p) - \frac{1}{2}\sigma_r(p)\right)\right|_{r=a} = \left.\sqrt{\frac{1}{4}\sigma_r^2(p) + \sigma_\theta^2(p) - \sigma_r(p)\sigma_\theta(p)}\right|_{r=a}$$
$$\leq \left.\sqrt{\sigma_r^2(p) + \sigma_\theta^2(p) - \sigma_r(p)\sigma_\theta(p)}\right|_{r=a} = \sigma_m(p) \tag{11}$$

Now we can refactor (8) with (11),

$$\sigma_m(\omega + p) \leq \sqrt{\sigma_m^2(\omega) + \sigma_m^2(p) + 2\sigma_m(\omega)\sigma_m(p)} = \sigma_m(\omega) + \sigma_m(p) \quad (12)$$

Based on this observation, the max Von-Mises stress of a flywheel is attributed to two factors:

$$\sigma_m \leq \sigma_m(\omega) + \sigma_m(p) \approx C_1\omega^2 + C_2 u' \quad (13)$$

Where $C_1$ and $C_2$ are constants related to flywheel geometry and material, and $u'$ is the shrink-fit percentage. In summary, if $\sigma_{vm}(\omega) + \sigma_{vm}(p)$ is within the material's strength. The flywheel will not fail. $\sigma_{vm}(p)$ and $\sigma_{vm}(\omega)$ can be evaluated separately based on rotational and external pressure, facilitating the design process. Figure 5 shows that the max Von-Mises is linear to both shrink-fit percentage and square of the rotational speed.

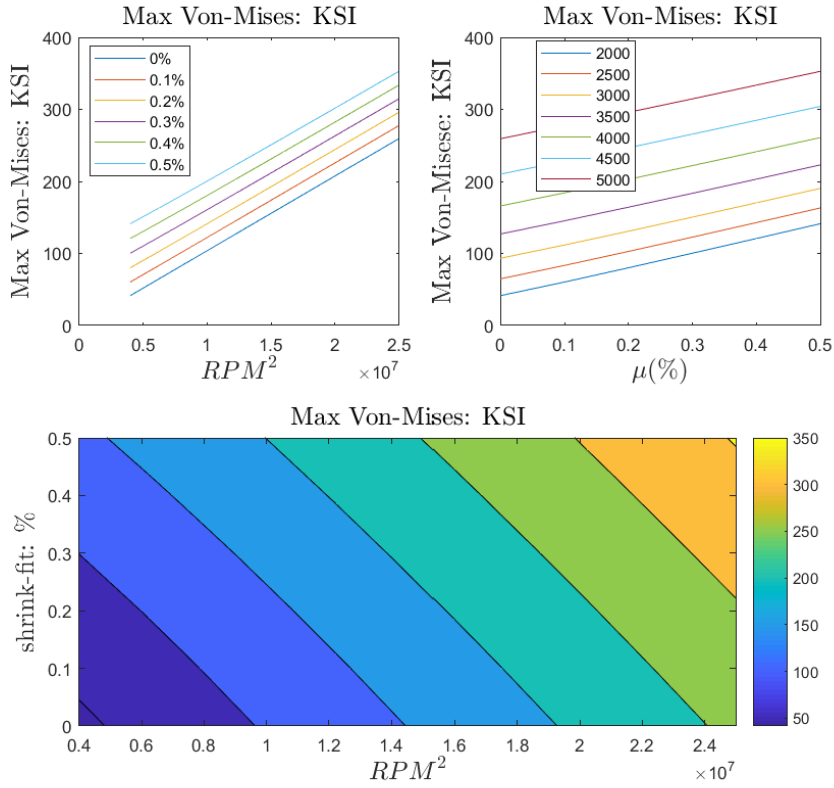

Figure 5 The max Von-Mises of a 2.13-meter annulus flywheel w.r.t the inner radius shrink-fit and rotational speed

The maximum radial, tangential stress occurs at its center for a solid, shaftless flywheel. The maximum Von-Mises stress also occurs at the flywheel's center, equal to $\frac{1}{8}(3 + v)\rho\omega^2 b^2$.

## Energy Density Analysis

This section uses the shaftless flywheel as a benchmark against two popular designs. The first case noted as Type I includes a borehole flywheel or rim shrink-fitted with a shaft, which acts as an interface to the magnetic bearings and motor/generators. The other case noted as Type II, utilizes an annulus flywheel without a shaft.

### *Annulus flywheel with a shrink-fitted shaft (Type I)*

This is a typical case where a shaft supports a flywheel. Most of the moment of inertia in the system is provided by the rotor/rim. The flywheel is subject to both rotational and inner pressure-caused stress. The combined mass of flywheel and shaft is larger than the mass of a solid flywheel with the same radius, which leads to lower specific energy. Furthermore, the specific energy is further lowered because the center hole confines the maximum spinning speed. The stress caused by interference fit ($\Delta\sigma$) is also detrimental to the flywheel.

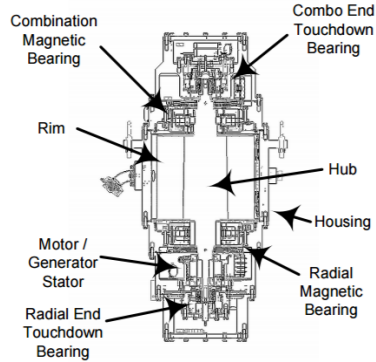

Figure 6 The NASA G2 flywheel[21]. It uses a conventional annulus design

Consider the max angular velocity is bounded by the stress $\sigma_u$ that the flywheel can undertake,

$$\omega^2(shaftless) = \frac{8\sigma_y}{(3+v)\rho b^2}$$
$$\omega^2(Type\ I) = \frac{8(\sigma_y - \sigma_s)}{(3+v)\rho b^2}\left(\frac{1}{\kappa}\right) \quad (14)$$

where $\kappa = \frac{2-2v}{3+v}t^2 + 2$. $t$ denotes the inner-to-outer radius ratio(equivalent to shaft-to-flywheel radius) for the annular flywheel ($t = a/b$). $\sigma_y$ is the material yield stress, $\sigma_s$ the shrink-fit caused stress. Based on the max angular velocity, we have the specific energy as

$$\frac{E}{m}(shaftless) = \frac{2\sigma_y}{\rho(3+v)}$$
$$\frac{E}{m}(Type\ I) = \frac{(t^2+1)(\sigma_y - \sigma_s)}{\rho[3+v+(1-v)(t^2)]}$$
(15)

The lift ratio of specific energy between a shaftless and annulus flywheel is

$$\lambda_I = \frac{E}{m}(shaftless) \div \frac{E}{m}(Type\ I)$$
$$= \frac{2}{(1-\Delta\sigma)}\left[1 - \frac{2(1+v)t^2}{(3+v)(t^2+1)}\right]$$
(16)

where $\Delta\sigma$ denotes the ratio of shrink-fit-caused stress to the material's tensile stress. Depending on the inner radius of the flywheel and the initial inference fit stress, the specific energy of an annulus flywheel with a shaft could be 50% or less than that of a solid shaftless flywheel.

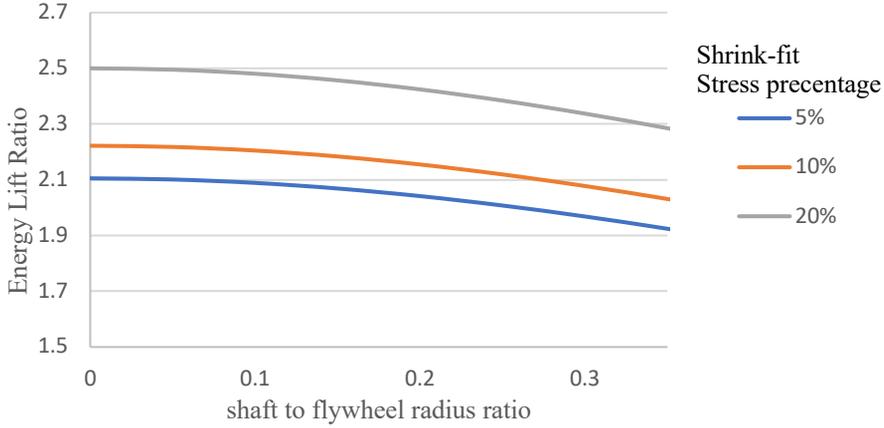

Figure 7 The ratios of specific energy and energy density of shaftless to annular flywheels.

For Figure 7, the horizontal axis is the ratio of shaft radius to flywheel radius. The different curves are the results of stresses caused by different shrink-fit allowances. In general, the shaftless flywheel will have doubled energy density.

## *Annulus steel flywheel without shaft (Shell flywheel, Type II)*

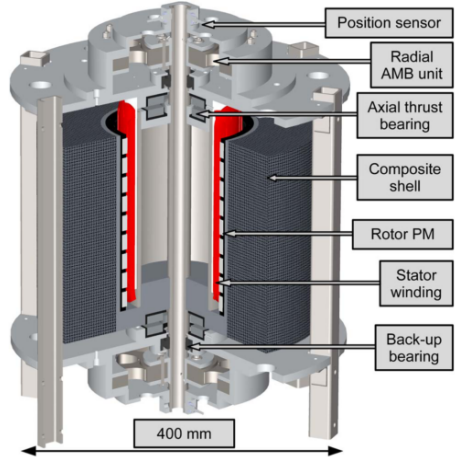

Figure 8 A shell flywheel[9] includes magnetically permeable materials as a part of the rim structure.

An annulus spinning disk's specific energy and energy density depends on its outer radius and rotational speed. This type of flywheel requires specialized magnetic bearing and motor design. The lift ratio between a shaftless and shell flywheel is:

$$\lambda_{II} = 2\left(\frac{\frac{1-v}{3+v}t^2 + 1}{1+t^2}\right) \tag{17}$$

$$\lim_{t\to 0}\lambda_{II} = 2, \qquad \lim_{t\to 1}\lambda_{II} \approx 1.2$$

Based on the inner and outer radius ration $t$, the lift ratio varies from 2 to 1.2, meaning the energy density of a thin wall flywheel could be 50% to 83% of a shaftless flywheel. However, 83% can only be achieved when $t = 1$, creating a thin-wall-like flywheel and significantly reducing its mass. Making such a flywheel requires a significant waste of materials.

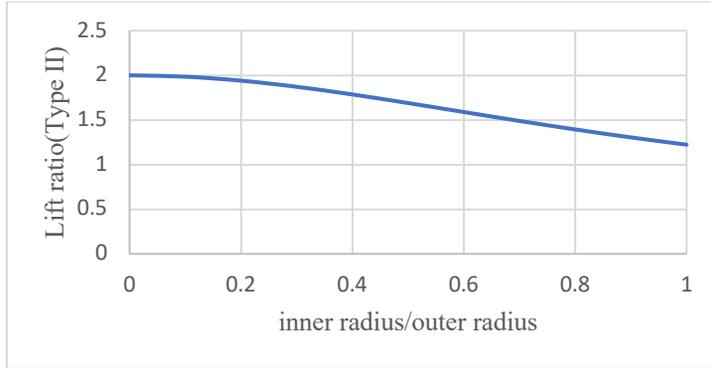

Figure 9 The lift ratio of specific energy between shaftless and shell flywheel with the same outer radius

## Shaftless Flywheel Design and Improvement

The shaftless flywheel is designed to facilitate magnetic levitation, motoring, and drop protection. Its section view is given in Figure 10. The upper body includes an extra rim to allow radial levitation fluxes. Its lower body has a large ring attached for drop protection. In addition, there is a slot for installing the motor magnets. The shaftless flywheel is an extension and physical realization of the concepts by the co-authors [22,23]. This effort was supported by the U.S. Department of Energy, Calnetix Inc and Texas A&M University. The magnetic bearing, motor design, and levitation control are detailed in [17,24,25].

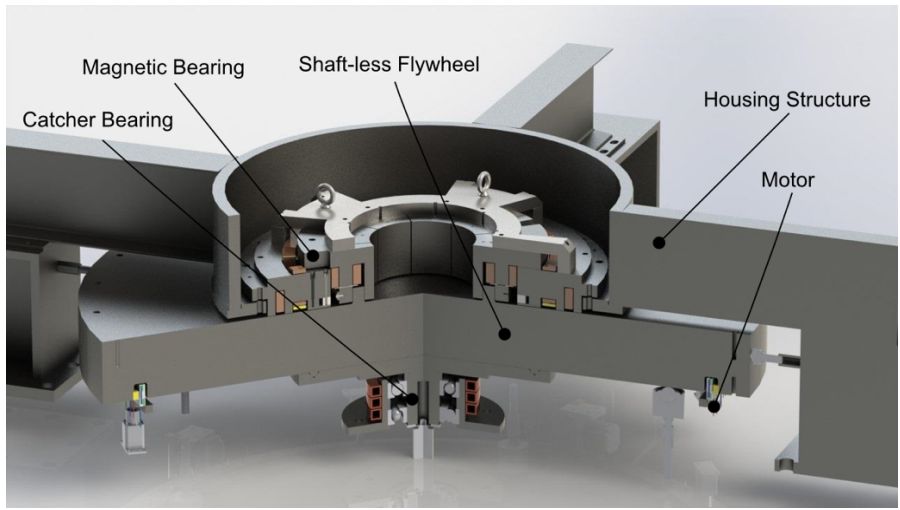

Figure 10  Section view of the shaftless FESS with system components illustrated

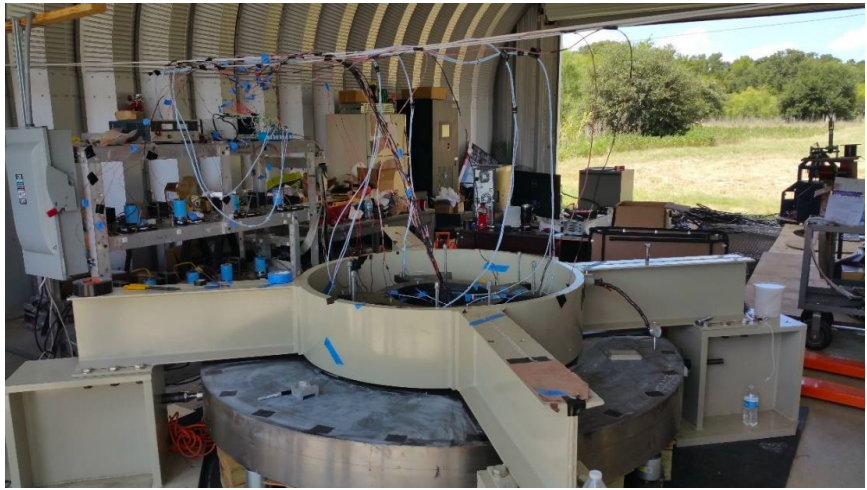

Figure 11  Full-scale shaftless flywheel

The nonlinear stress analysis was carried out using commercial finite element analysis software. It was found that, for the existing design, the max stress occurs around the flywheel's radial ring and the main body. It is below the heat-treated 4340 stainless steel yielding strength if the flywheel operates below 4500 RPM. However, as depicted in Figure 12, higher spinning speed will cause the flywheel to fail.

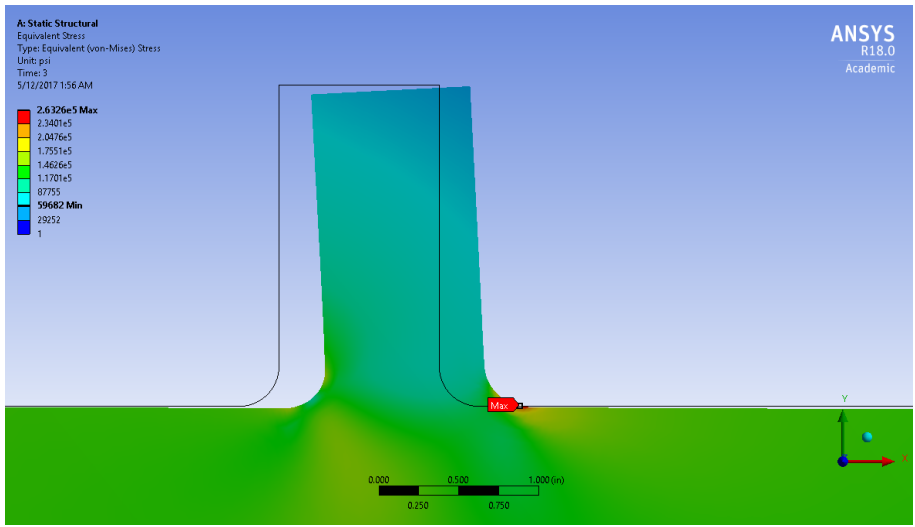

Figure 12  A close look at the stress concentration around the outer radius fillet at a higher speed (before optimization). The deformation was amplified by 10, notice the max V.M. stress is close to 265 ksi (1820 MPa).

In the following, an amendment to the current design is proposed for higher spinning speed and energy density. Since the stress concentration at the outer fillet is tensile, it can be countered by compressive preload created by shrink-fitting an extra inner ring to the radial rim. In order to achieve the best results, the shrink-fit percentage and other design parameters, including the fillet radius ($R_r$), radial ring height ($H_r$), radial ring thickness ($Th_r$) and motor slot fillet ($R_m$), are optimized to reduce the maximum stress of the entire flywheel. In addition, to ensure the new flywheel will fit the existing magnetic bearing system, the shrink-fit ring thickness ($Th_s$) is constrained to be ($25\text{mm} - Th_r$), where 25mm is the magnetic pole size of the present flywheel. An optimization process is carried out to find the proper values for the design parameters.

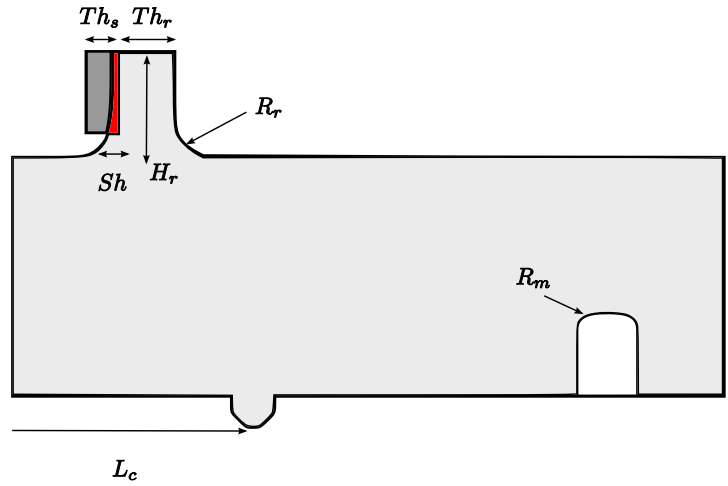

Figure 13  Flywheel design modification and the parameters to be optimized (Dimension not drawn to scale)

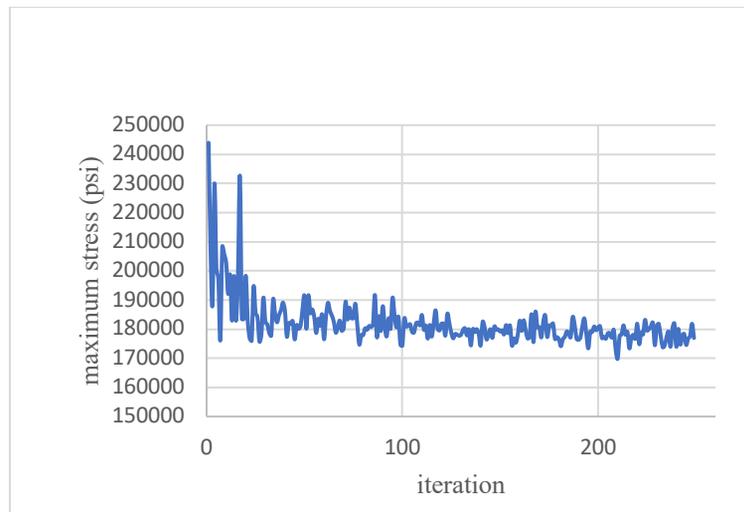

Figure 14  Simulation results show the max stress being reduced by optimizing

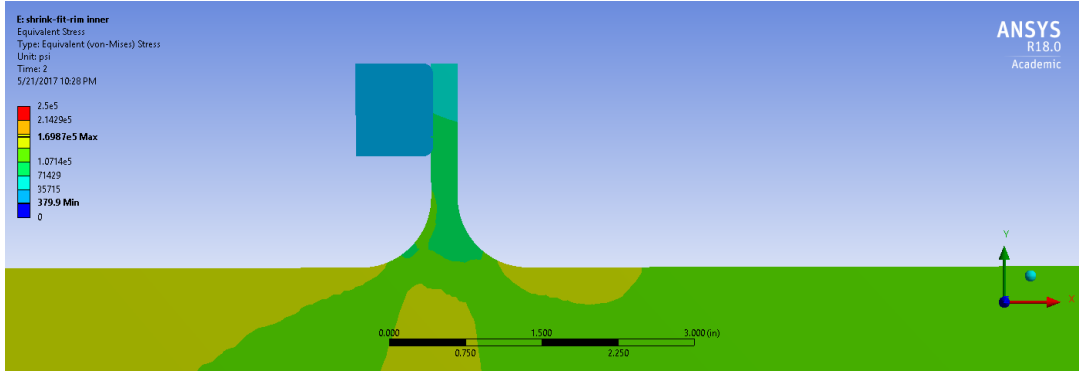

Figure 15   A close look at the improved stress distribution around the radial ring. The max V.M stress is close to 180 ksi (1240 MPa)

The optimization process utilizes an adaptive single target solver. At first, the parameters are treated as continuous variables rather than manufacture-feasible variables to reduce complexity. When a satisfying result is achieved, the design parameters are adjusted to manufacture-possible values, and simulation results are verified again. The resulting improvements are summarized in the following table. The new flywheel design will achieve specific energy of 23 Wh/kg and an energy density of 185 KWh/m$^3$.

TABLE II Improved Flywheel Specifications

| Parameter Name | Quantity | Unit | Improvement To initial design |
|---|---|---|---|
| Mass | 5443 | [kg] | - |
| Moment of inertia | 3087 | [kg·m$^2$] | - |
| Max rotational speed | 5623 | [rpm] | 12.46% |
| Tip speed | 588.89 | [m/s] | 12.46% |
| Max Energy | 148 | [kWh] | 26.5% |
| Operational Energy | 126 | [kWh] | 26.0% |
| Power capacity | 100 | [kw] | - |